\documentclass{webofc}
\usepackage[varg]{txfonts}   

\begin{document}
\title{Particle Flow and PUPPI in the Level-1 Trigger at CMS for the HL-LHC}

\author{\firstname{Benjamin} \lastname{Kreis}\inst{1}\fnsep\thanks{\email{kreis@fnal.gov}} 
        \firstname{} \lastname{for the CMS Collaboration} 
}

\institute{Fermi National Accelerator Laboratory, Batavia, IL 60510, USA
          }

\abstract{%
With the planned addition of tracking information to the Compact Muon Solenoid (CMS) Level-1 trigger for the High-Luminosity Large Hadron Collider (HL-LHC), the trigger algorithms can be completely reconceptualized.  We explore the feasibility of using particle flow-like reconstruction and pileup per particle identification (PUPPI) pileup mitigation at the hardware trigger level. This represents a new type of multi-subdetector pattern recognition challenge for the HL-LHC. We present proof-of-principle studies on both physics and hardware-resource performance of a prototype algorithm for use by CMS in the HL-LHC era.
}
\maketitle
\section{Introduction}
\label{intro}


In the mid-2020s, the High-Luminosity Large Hadron Collider (HL-LHC) at CERN will begin colliding protons with an instantaneous luminosity leveled to $5 \times 10^{34} \ \mathrm{cm}^{-2}\mathrm{s}^{-1}$ with the goal of delivering approximately $300 \ \mathrm{fb}^{-1}$ per year for ten years \cite{hllhc}.
This increase in luminosity will enable the general-purpose experiments at the HL-LHC, CMS and ATLAS, to maintain progress in their rich physics programs 
of precision measurements of the standard model and searches for physics beyond the standard model, such as supersymmetry and dark matter.  

The challenge for the HL-LHC experiments will be to maintain detection efficiency for interesting physics processes at the electroweak energy scale 
occurring in the same bunch crossings as additional inelastic interactions, known as {\it pileup}.  
Pileup produces extra hits in the tracking subdetectors and extra energy depositions in the calorimeter subdetectors that must be separated from the process of interest.
In addition, the subdetectors must be able to withstand the radiation damage caused by the particles produced by these many proton interactions.
At the HL-LHC, the average number of pileup interactions is expected to reach 200.

The CMS experiment \cite{cms} will undergo a comprehensive upgrade to maintain performance in the high-luminosity conditions of the HL-LHC.
In this paper, we focus on upgrades to the Level-1 trigger, the first stage of data processing that has a coarse, global view of the detector and determines 
which collision events to fully read out for further analysis. 
We explore the feasibility of using particle-level reconstruction and pileup mitigation, the best performing algorithms in current offline processing, in the upgraded trigger.
In Section \ref{sec:level1}, we describe the inputs to the upgraded Level-1 trigger and its architecture.  
The inputs will include tracks for the first time, and the trigger will have a {\it correlator} that forms the global view of each collision event.  
Particle-level reconstruction and pileup mitigation algorithms for the correlator are described in Section \ref{sec:algo}.  
In Section \ref{sec:res}, results on physics performance and hardware resource requirements based on a proof-of-principle study are presented.

\section{Level-1 Trigger for HL-LHC}
\label{sec:level1}

Upgrades to the CMS detector for HL-LHC, described in detail elsewhere \cite{cmsupgrade,interim}, will bring new and improved inputs to the Level-1 trigger.
The inputs, referred to as {\it trigger primitives}, are formed by trigger primitive generators (TPGs).
For the first time at CMS, the Level-1 trigger will receive trigger primitives representing tracks 
with transverse momentum $p_T$ $>$ 2-3 GeV
from a new outer tracking subdetector.
The TPGs of a new endcap calorimeter will provide energy deposition information at improved granularity.  
New TPGs in the barrel electromagetic calorimeter will also improve granularity,
while the TPGs of the hadron calorimeter will be upgraded without changing granularity.
Finally, the TPGs of the muon subdetectors will upgraded and will include trigger primitives from new muon detectors added to the endcap region.

\begin{figure*}[tbh!]
\centering
\includegraphics[width=12cm]{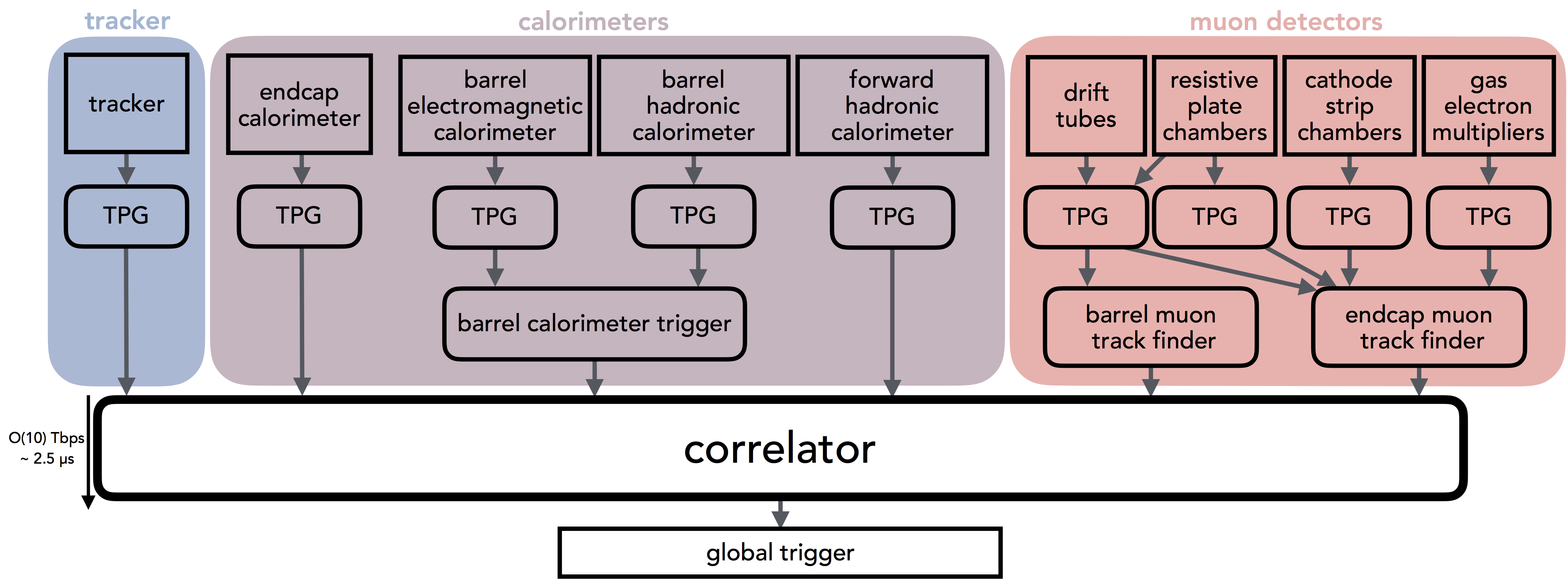}
\caption{Schematic of the CMS Level-1 Trigger for HL-LHC.  A number of possible, additional direct links are not pictured (see \cite{interim}). }
\label{fig:l1t}
\end{figure*}

The data flow from the CMS subdetectors to the TPGs and through the Level-1 trigger is shown in Figure \ref{fig:l1t}.  
The barrel calorimeter trigger processes trigger primitives to form high-resolution clusters.
The barrel and endcap muon track finders identify muons.
The output of the barrel calorimeter trigger, muon track finders, and the trigger primitives of the tracking subdetector, endcap calorimeter, and forward hadronic calorimeter
reach the correlator at a bandwidth of  $\mathcal{O}(10)$ Tbps, where the information is correlated to form a global view of the event.  
This is passed on to the global trigger, where the trigger decision is made.

The Level-1 trigger systems will use Field Programmable Gate Arrays (FPGAs) for all data processing.  
Modern high performance FPGAs have high speed input/output able to handle the large data bandwidth 
and hundreds of thousands of logic cells to do pipelined and parallelized data processing.
The correlator has a latency budget of approximately 2.5 $\mu s$ and must be pipelined to the LHC collision rate of 40 MHz.

\section{Particle-Level Algorithms for Level-1 Trigger}
\label{sec:algo}

The availability of tracking information in the Level-1 trigger correlator opens up the possibility of 
performing reconstruction and pileup mitigation at the particle level, the best performing approach in CMS's software-level trigger and offline reconstruction \cite{pf,puppicms}.
To perform these algorithms in the correlator, they must be implemented in FPGAs and meet the pipelining and latency constraints.

\subsection{Particle-Flow Reconstruction in Level-1 Trigger}
\label{sec:pf}

\begin{figure*}[tbh!]
\centering
\includegraphics[width=13cm]{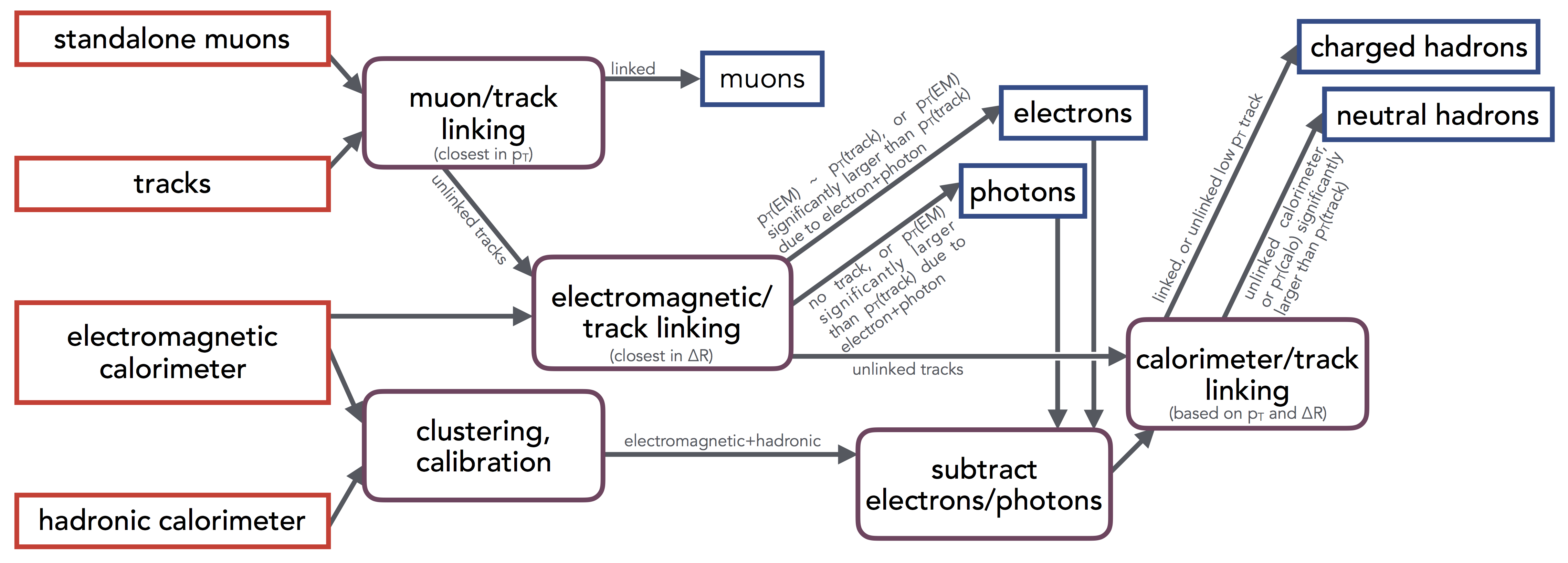}
\caption{Schematic of particle flow algorithm for CMS Level-1 trigger correlator.}
\label{fig:pf}
\end{figure*}

Particle-flow reconstruction correlates tracks from the tracking and muon subdetectors and clusters from the calorimeter subdetectors to identify each final-state particle 
and combines the subdetector measurements to reconstruct the identified particles' properties \cite{pf}.
Here we describe a proof-of-principle algorithm for performing particle-flow reconstruction in the FPGAs of the Level-1 trigger correlator.
A schematic of the algorithm is shown in Figure \ref{fig:pf}.

The algorithm inputs are muon candidates from the muon track finders, tracks from the tracking subdetector TPGs, 
and electromagnetic and hadronic calorimeter energy deposits from the barrel calorimeter trigger and endcap calorimeter TPGs.
The algorithm reconstructs particles centered within a restricted region in pseudorapidity ($\eta$) and azimuthal angle ($\phi$) based on inputs from an enlarged region to correlate across the $\eta$-$\phi$ boundaries.
Multiple copies of the algorithm will run in parallel to cover the whole detector. 

Muon candidates and tracks closest in $p_T$
within a $\Delta R \equiv \sqrt{\Delta \eta^2 + \Delta \phi ^2}$ requirement 
are linked to form particle-flow muon candidates.  
The linked tracks are removed from further consideration.  
Electromagnetic calorimeter clusters and tracks closest in $\Delta R$
within a $\Delta R$ requirement
are linked.  
Linked electromagnetic energy clusters and tracks form particle-flow electrons,
and the linked tracks are removed from further consideration.
If the electromagnetic energy cluster's $p_T$ is significantly larger than that of the track,
the excess $p_T$ forms a particle-flow photon.
Particle-flow photons are also formed from unlinked electromagnetic energy clusters.
The electromagnetic and hadronic calorimeter energy is clustered and calibrated, 
and the energy of the particle-flow electrons and photons is removed.
The resulting calorimeter clusters are linked with tracks based on $\Delta p_T$ and $\Delta R$.
Linked calorimeter clusters and tracks and unlinked low-$p_T$ tracks form particle-flow charged hadrons.
If the calorimeter cluster's $p_T$ is significantly larger than that of the track,
the excess $p_T$ forms a particle-flow neutral hadron.
Particle-flow neutral hadrons are also formed from unlinked calorimeter clusters.

\subsection{PUPPI Pileup Mitigation in Level-1 Trigger}
\label{sec:puppi}

\begin{figure*}[tbh!]
\centering
\includegraphics[width=10cm]{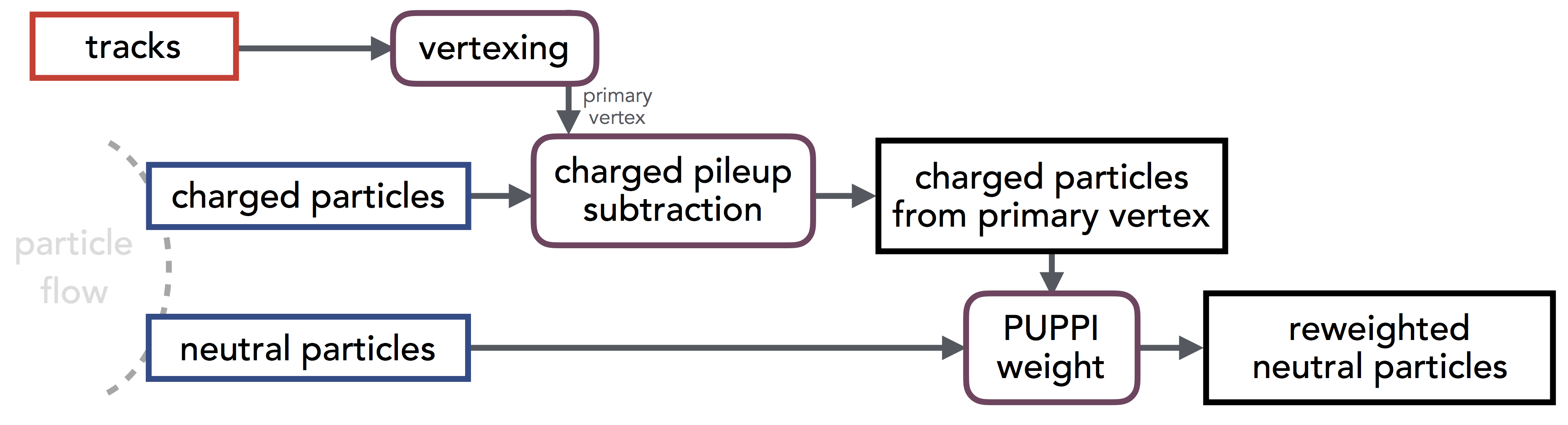}
\caption{Schematic of PUPPI pileup mitigation algorithm for CMS Level-1 trigger correlator.}
\label{fig:puppi}
\end{figure*}

The best performing pileup mitigation technique used in CMS offline reconstruction 
is PileUp Per Particle Identification (PUPPI), 
an algorithm that removes charged particles with tracks not originating at the primary vertex
and downweights neutral particles based on the probability that they originate from pileup \cite{puppi,puppicms}.
The weight is calculated from the neighboring charged particles not originating from pileup.
Here we describe a proof-of-principle implementation of PUPPI for the FPGAs of the Level-1 trigger correlator.
A schematic is shown in Figure \ref{fig:puppi}.

The position along the beam axis ($z$) of the primary vertex is computed from all tracks in the event.
One algorithm for finding the primary vertex is to find the $z$ position of the peak in a histogram of track $z$
weighted by track $p_T$.
Charged particles with tracks not originating near the primary vertex are removed.
For each neutral particle $i$, we compute the sum of $p_T / \Delta R_{ij}$ 
of the remaining charged particles $j$ within a $\Delta R$ requirement.  
The sum addresses a precomputed lookup table of weights.

\section{Results}
\label{sec:res}

\subsection{Physics Performance}
\label{sec:phys}

We test the physics performance of particle-flow reconstruction and PUPPI pileup mitigation in the Level-1 trigger
using simulated samples of $t\bar{t}$ events decaying semileptonically as signal and minimum-bias events as background.
The average number of pileup interactions in the simulation is 140, corresponding to a baseline HL-LHC scenario.  

\begin{figure}[tbh!]
\begin{center}
\includegraphics[width=0.45\linewidth]{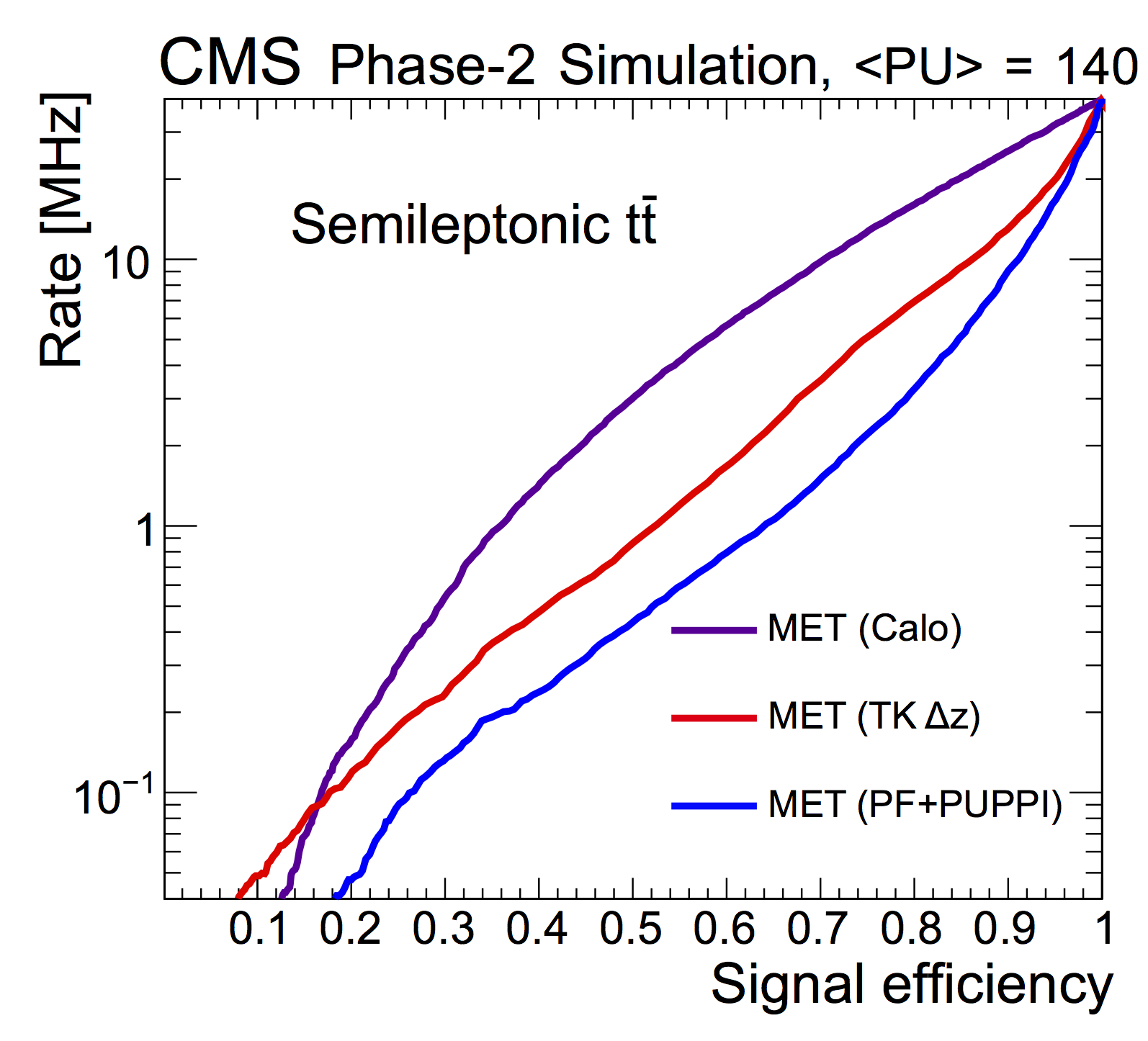}
\hspace{1cm}
\includegraphics[width=0.45\linewidth]{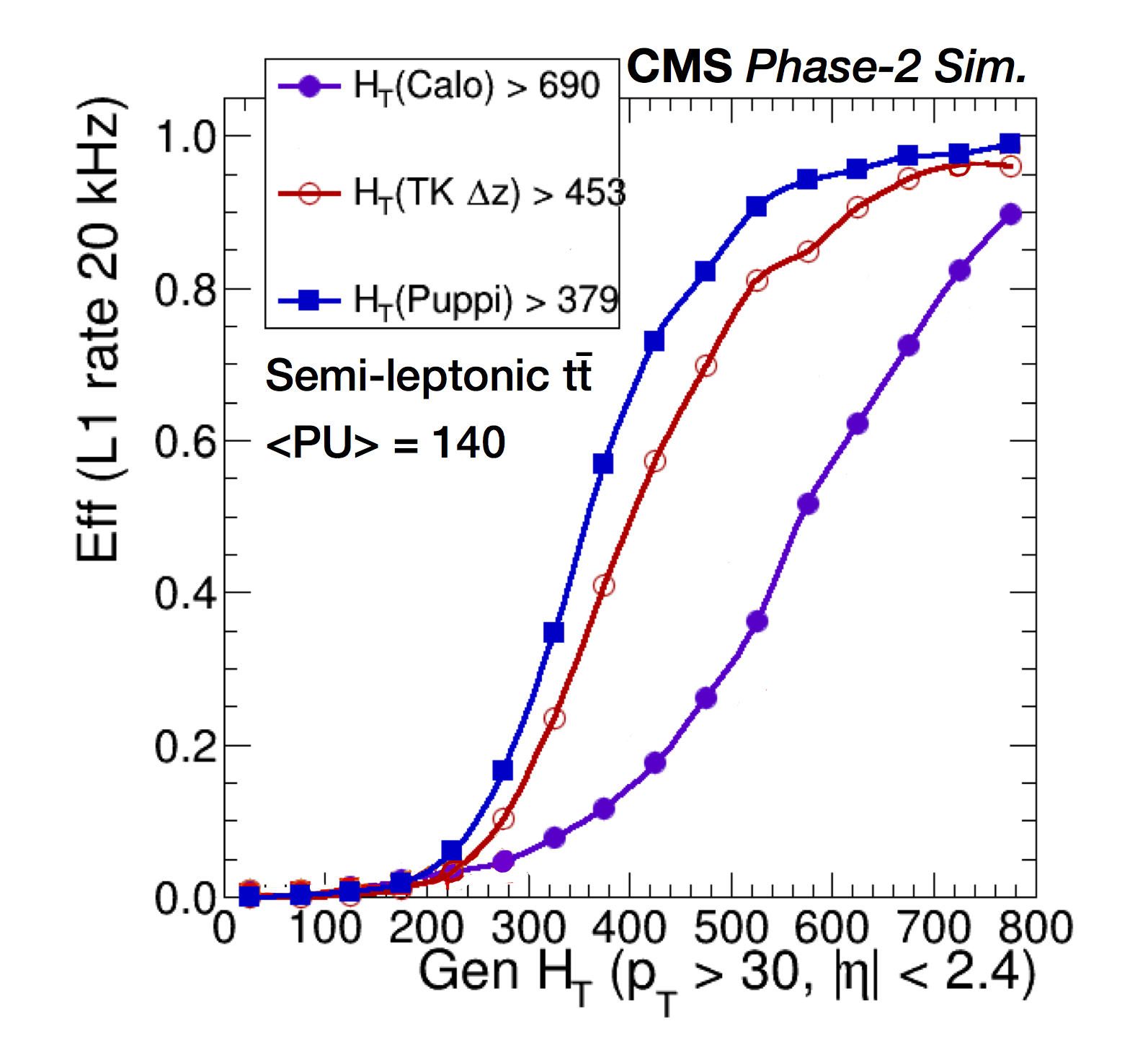}
\end{center}
\caption{MET trigger rate versus signal efficiency (left) and $H_T$ trigger turn-on curve (right)
for the particle-flow and PUPPI based trigger (blue), a calorimeter-only trigger (purple),
and a trigger based on tracks consistent with the primary vertex (red) \cite{interim}. }
\label{fig:phys}
\end{figure}

Figure \ref{fig:phys} (left) shows the rate of a trigger on missing transverse energy (MET) computed with particle-flow candidates 
versus signal efficiency for events with MET>100 GeV at the generator level.  
We compare with two alternative MET triggers.  The first computes MET from calorimeter energy deposits.  
The second computes MET from tracks with a small $\Delta z$ with respect to the primary vertex.  
Figure \ref{fig:phys} (right) compares trigger efficiency at a fixed rate of 20 kHz versus the scalar sum of jet $p_T$ ($H_T$) at the generator level.  
The jets are clustered using the anti-$k_T$ algorithm with a radius of 0.4 \cite{antikt} and are required to have $p_T>30$ GeV and $\eta<2.4$.


%
%
%

\subsection{Hardware Implementation}
\label{sec:hw}

We divide the detector into $\eta \times \phi \approx 0.55 \times 0.55$ regions accepting up to 25 tracks and 20 calorimeter clusters each.  Implmented in Xilinx Vivado High Level Synthesis version 2016.4 for an Ultrascale+ VU9P FPGA, four regions can be processed with approximately $40 \%$ of the FPGA resources.  The latencies for particle flow and PUPPI are approximately 500 ns and 100 ns, respectively.

\section{Conclusions and Outlook}
\label{sec:conc}


We explore the feasibility of using particle-flow reconstruction and PUPPI pileup mitigation, 
the best performing algorithms in current offline processing, in the CMS Level-1 trigger correlator
to maintain performance in the high pileup conditions of the HL-LHC.
We present proof-of-principle implementations of these algorithms for FPGAs
and find that they significantly improve the physics performance of the Level-1 trigger
while being feasible in terms of FPGA resource usage and latency.

The concept of adapting algorithms used in offline processing to the FPGAs of the Level-1 trigger may give rise to
additional opportunities to improve physics performance.  
We note in particular that machine learning techniques, which are being explored for offline processing,
are also a promising direction for the Level-1 trigger \cite{hls4ml}.

\end{document}